\documentstyle[12pt,epsfig]{article}

\def\be{\begin{equation}}
\def\ee{\end{equation}}
\def\bear{\begin{eqnarray}}
\def\eear{\end{eqnarray}}
\newcommand{\beq}{\begin{eqalignno}}  
\newcommand{\eeq}{\end{eqalignno}}
\def\benu{\begin{enumerate}}
\def\eenu{\end{enumerate}}
\def\lt{<}
\def\ncom{\newcommand}
\ncom{\mbf}[1]{\mathbf{#1}}
\ncom{\mca}[1]{\mathcal{#1}}

\begin{document}
%

\vskip -1cm
\begin{flushright}
{
UM-TH-02-07}
\end{flushright}
\vspace*{0.5cm}

\begin{center}


{\Large{\bf Supersymmetry and the Cosmic Ray Positron Excess } }

\vspace{1cm}
{\large G.L. Kane\footnote{gkane@umich.edu}, Lian-Tao Wang\footnote{liantaow@umich.edu} 
and Ting T. Wang\footnote{email: tingwang@umich.edu}} \\
\vspace{0.15cm}
{\it  Michigan Center for Theoretical Physics, 
University of Michigan, Ann Arbor, MI-48109 }




\begin{abstract}
We explore several supersymmetric alternatives to explain  
predictions for the cosmic ray positron excess. Light sneutrino or
neutralino LSP's, and a fine-tuned model designed to provide a
$\delta$-function input, can give adequate statistical descriptions of
the reported HEAT data if non-thermal production of the relic cold dark
matter density dominates and/or if ``boost factors''(that could originate in
uncertainties from propagation or local density fluctuations)  to
increase the size of the
signal are included. All the descriptions can be tested at the Tevatron
or LHC, and some in other WIMP detecting experiments. 
\end{abstract}
\end{center}


\section{Introduction}
The recent HEAT experiment \cite{heat:2000} has confirmed an excess and possible 
structure around $8$ GeV in the positron spectrum of cosmic rays \cite{Barwick:1997ig}\cite{Barwick:1995gv}. 
Since it appears to be  unlikely that conventional astrophysical
phenomena can lead
to such structure, and since it is 
plausible that the annihilation of WIMPs  in the galactic halo can 
give rise to a high energy positron excess, this 
unexpected feature of the positron spectrum could be a major discovery. LSPs
(the lightest supersymmetric particles) are stable particles predicted by  
supersymmetric standard model. They are natural candidates for the cold dark 
matter which forms the galactic halo, and thus could be the needed
WIMPs.  Therefore, it is interesting to
examine in detail whether LSPs can quantitatively explain the
observations. 

The conventional candidate for the LSP has been the lightest neutralino. 
The resulting positron spectrum from the annihilation of those neutralinos 
was studied \cite{Turner:1989kg, Tylka:xj, Kamionkowski:1990ty,
Baltz:1998xv}  and re-examined  after the newest HEAT result was
announced \cite{Kane:2001fz, Baltz:2001ir}. The general 
result has been 
\begin{enumerate}
\item If the mass of the LSP is less than the W mass, the cross section for  
a pair of neutralinos to annihilate into a pair of fermions tends to
be very small due to the well known suppression proportional to the
fermion masses.  In this case the positron excess is far too small. 

\item If the mass of the LSP is larger than the W mass (but smaller than
the top mass), the annihilation 
to a pair of W's always dominates. The positrons coming from the direct 
decay mode $W^+ \rightarrow e^+ + \nu $ will produce a large excess at
and below an energy of about half the W mass. Additional positons come from
secondary decays of b, $\mu\,,\tau$, etc. That excess, after 
propagation, can be extended substantially to lower energies. While it
is hard to reproduce the apparent energy dependence of the 
HEAT data, it is possible to have an excess in the observed region. 

\item The actual 
positron flux resulting from the annihilation is always too small to 
produce visible structure. The positron signal in the literature has
been increased by a 
``boost factor'' which is sometimes large. This boost factor may 
be explainable by the uncertainties in the propagation process or the 
clumpyness nature of the galactic halo \cite{Sikivie:1995dp}. The need
for such a
factor means the HEAT data alone cannot guarantee that superpartners and
LSP cold dark matter have been observed, though if confirmed and not
explained by conventional cosmic ray processes the superpartner
discovery would be a favored option.  At the same time, a need for
a medium or large ``boost factor'' should not be viewed as excluding a
particular LSP since large boost factors could actually be physical.
\end{enumerate}

\begin{figure}
\epsfig{file=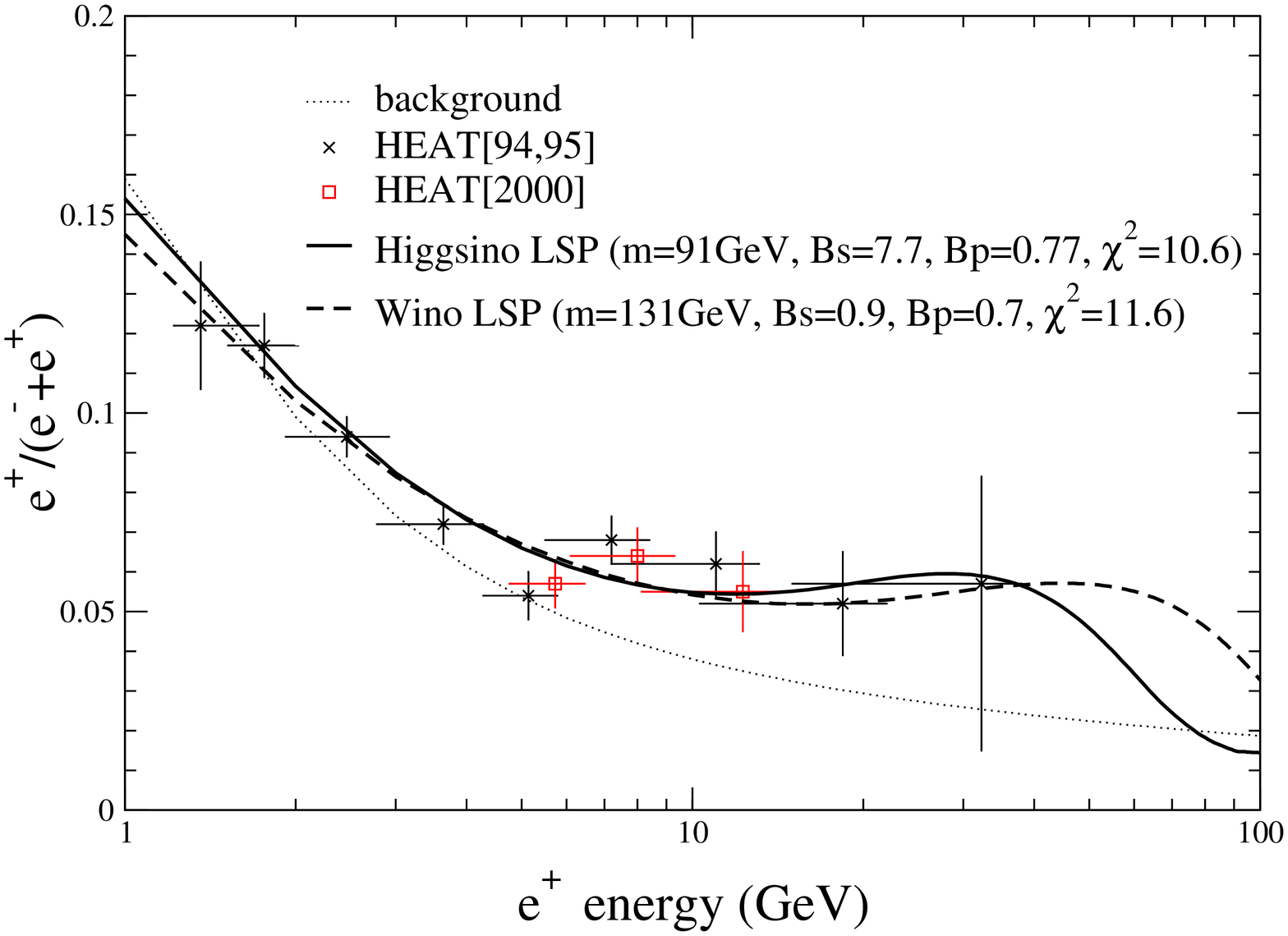,height=15cm,angle=-90}
\bigskip
\caption{Numerical results from   different neutralino LSP models. The
relic density is assumed to arise from non-thermal mechanisms and is
normalized to the average local cold dark matter density.  Bs is the boost
factor defined in the text, possibly due to local dark matter
concentrations and to uncertainties in propagation, and Bp  the normalization factor multiplying
the background from \cite{Moskalenko:1997gh}. The $\chi^2$ values  are given to help
judge the quality of the description of the data.  The two cases are
for mainly Higgsino and  mainly wino LSP's;
mainly bino  cases need much more enhancement. These examples are for
neutralinos that are  consistent with all collider and direct detection
data. The Higgsino-like  case $M_1=500$ GeV, $M_2= 400$ GeV, $\mu = -100$
GeV, $\tan \beta = 10 $. The wino-like case $M_1=500$ GeV, $M_2= 165$ GeV, $\mu = 225$
GeV, $\tan \beta = 10$. }
\label{spectrum1}
\end{figure}

We use DARKSUSY \cite{Gondolo:2000ee} for calculating the positron flux.
The results for the neutralino case are shown in
Fig~\ref{spectrum1} for particular mass and type of LSP's. Mainly bino
LSP's do not work since they do not annihilate to W's.   The results
are not too sensitive to
the mass once it is above $m_W$, but extra kinetic energy for the W's
will spread out the
spectrum.  We use the formula for background
positrons produced by conventional sources provided by
\cite{Moskalenko:1997gh}  and also used by\cite{Baltz:1998xv,
Baltz:2001ir}. We also
treat the 
overall normalization of the background as a free parameter of
${\mathcal{O}}(1)$ and include it 
in the combined fit performed here, since the background
parameterizations were previously done assuming no new physics signal.  In
this and all the examples
considered below, the relic density is normalized to a local density
$\rho=0.3$ GeV${/cm^3}$.   

\begin{figure}
\epsfig{file=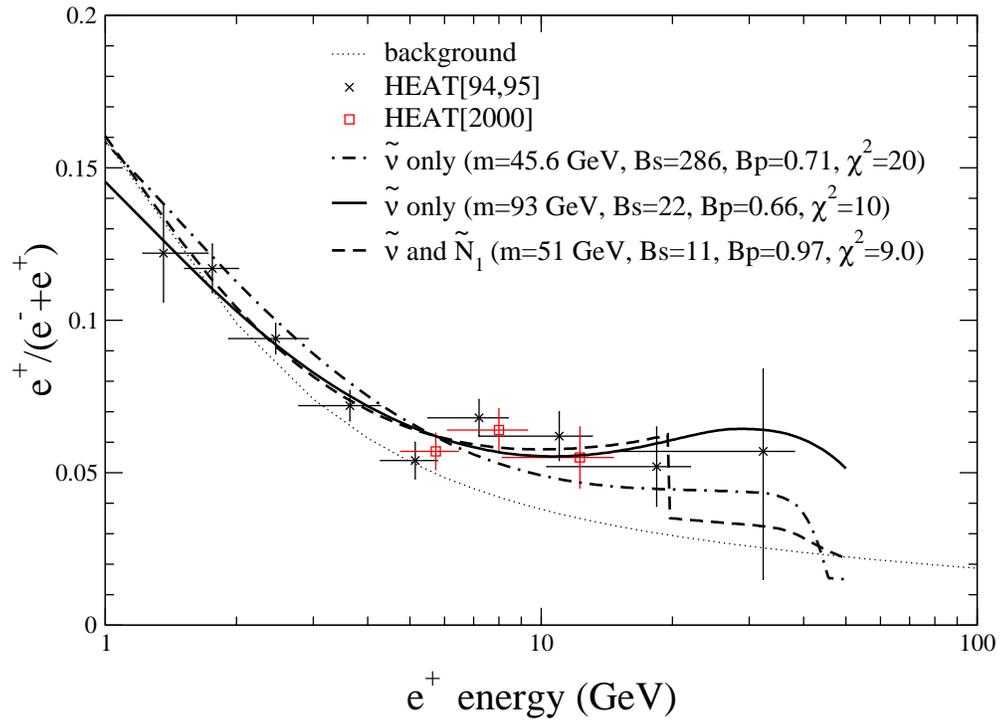,height=15cm,angle=-90}
\bigskip
\caption{Numerical results from alternative LSP models. See Fig 1 caption. 
The cases displayed here are very light sneutrino ($m_{\tilde{\nu}}
\sim \frac{1}{2} m_Z)$, sneutrino LSP $m_{\tilde{\nu}}> m_{W}$, and
sneutrino-neutralino degenerate models. All models are consistent with
collider data. }.  
\label{spectrum2}
\end{figure}

Since the conventional scenario does not give a decisive answer, 
it is interesting to explore alternative scenarios of the 
nature of the LSP and dark matter in order to see if any provide better
results. We explore 
sneutrino LSPs in Fig~\ref{spectrum2}, which leads to results similar to the 
conventional scenario if $m_{\tilde{\nu}} \stackrel{>}{_{\sim}} m
_W$. 
Since the HEAT data suggests structure in the energy spectrum, we are even willing 
to try models with some extreme ideas. The best case we could come up with 
is a sneutrino and a bino-like neutralino which are degenerate in 
mass.
Such a model does give more structure around $8$ GeV, also shown in
Fig~\ref{spectrum2}.  We also check the 
two scenarios against all available experimental constraints, mainly
from LEP,  and 
examine their implications for discoveries at the Tevatron. The
sneutrino examples would be excluded by the absence of observation in
direct detection experiments, but such exclusions are model
dependent. For example, the models of ref.\cite{Hall:1997ah,
Arkani-Hamed:2000kj,Smith:2001hy} evade such constraints. Similarly, the limits
could be evaded by right-handed sneutrinos. 

\begin{figure}
\epsfig{file=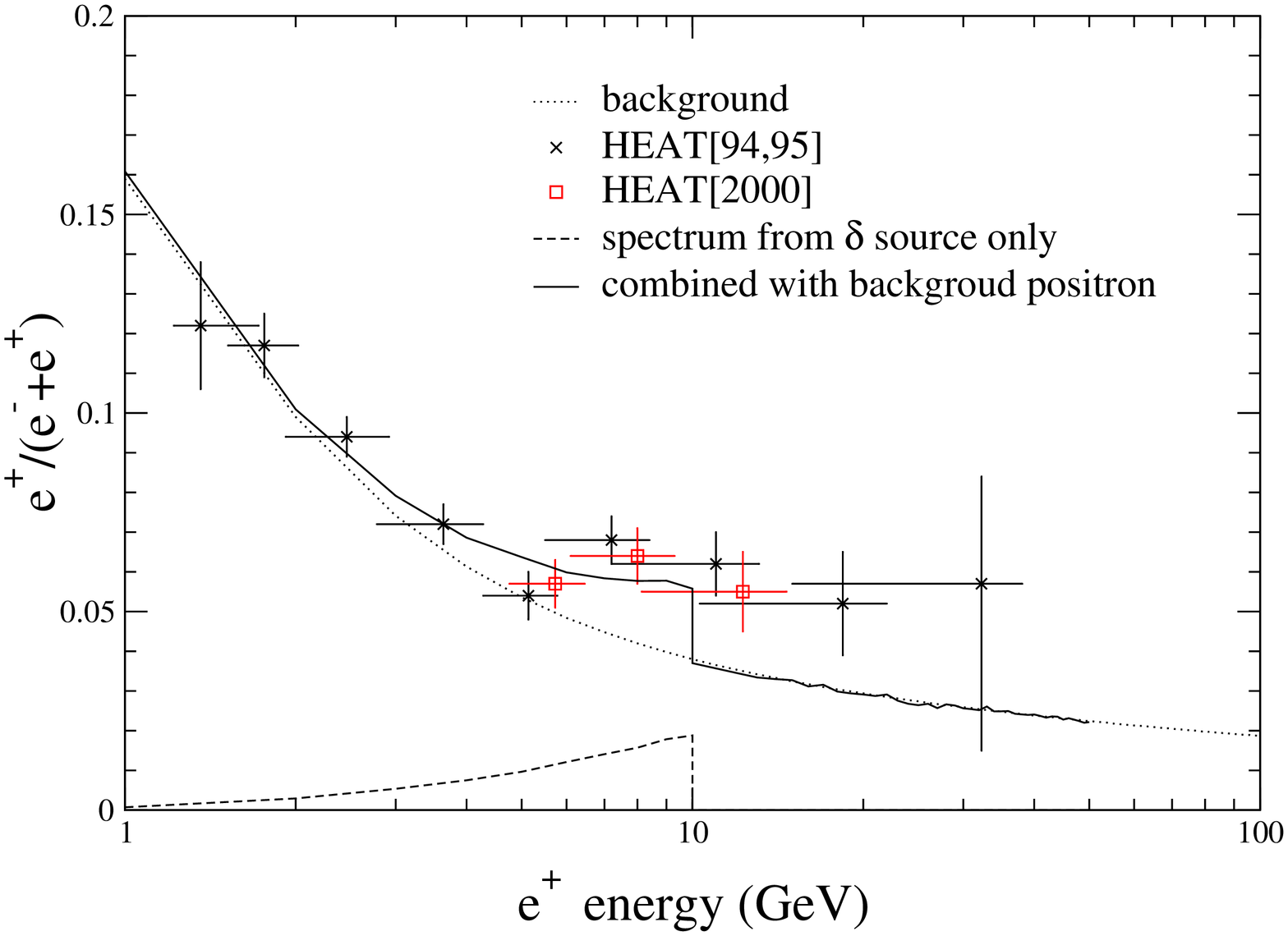,height=15cm,angle=-90}
\bigskip
\caption{A $\delta$-function source at $10$ GeV will be spread out by
propagation  at
lower energies as in the dashed curve, when added to the positron
background it can give the solid line. The position of the
$\delta$-function can be moved to higher energies -- this figure is only
to illustrate that such a result is the best one possible.}
\label{delta}
\end{figure}

It should be noted that it is very hard to obtain a 
``bump'' like structure from any positron production mechanism. The best
case would be a $\delta$ function as the initial positron energy
distribution, but the energy loss  
will only extend the distribution toward the low energy direction rather 
than spread it out, as shown in Fig~\ref{delta} 

Our results for neutralinos differ somewhat from those of Baltz et
al.\cite{Baltz:2001ir}  since we do not force thermal equilibrium relic densities
to account for the dark matter. We allow non-thermal mechanisms
\cite{Kallosh:1999jj,Giudice:1999yt,Giudice:1999am,Gherghetta:1999sw,
Jeannerot:1999yn,Moroi:1999zb}  to
dominate, and normalize to average local relic density. Similarly,
some of our examples will imply $\bar{p}$ rates somewhat above current
measuments\cite{Orito:1999re,Maeno:2000qx,Beach:2001ub} if naive $\bar{p}$
propagation were  correct, but most people
feel there are large uncertainties that mean one should not take such
results overly seriously. Also, the $\bar{p}$ rate is somewhat
sensitive to particle physics details, particularly the neutralino
annihilation mechanisms. Thus while the $\bar{p}$ rate should be kept
in mind we do not think it is a  compelling constraint at the present time. 

\section{Light $\tilde{\nu}$ LSP}
For a sneutrino with mass larger than W boson, the annihilation to a
pair of W's will dominate. The resulting positron spectrum will 
be similar to the one we get from neutralino annihilation. Recent
studies \cite{Altarelli:2001wx, Cho:1999km} show that a lighter sneutrino, close to the LEP
lower bound, is actually favored by the electroweak precision
data. Therefore, we also explore the scenario in which a light sneutrino with
a mass just above the LEP lower limit is the LSP. 

Similar to the neutralino annihilation, the annihilation of a pair of
sneutrinos to a pair of fermions is still
suppressed by the fermion masses or by the p-wave scattering due to
angular momentum conservation. However, if $m_{\tilde{\nu}} \approx m_Z/2$ the s-channel
annihilation is enhanced significantly by the Z-pole contribution, leading to a larger excess of positrons. However, it turns
out that this is still not sufficient to produce an excess as large as
that resulting from W decays. Consequently a very large boost factor is
needed.                                    

\subsection{Numerical Result}

If $m_{\tilde{\nu}} \lt m_W$, the s-channel production through the Z-pole
gives the dominant contribution and t-channel exchange of a chargino is
negligible. If $m_{\tilde{\nu}} \stackrel{>}{_{\sim}} m_W$, the most
important  processes for
the W pair production is the 4-point vertex interaction and s-channel
higgs boson exchange.  We use CompHEP \cite{Pukhov:1999gg}  for calculating the cross
sections. We  included a boost factor in the combined fit as described
above.  Both cases of sneutrino masses are studied. We also assume here the
dark matter only has one generation of sneutrinos, and sneutrino and
anti-sneutrino have the same number density.  The results are shown in 
Fig~\ref{spectrum2}.  

The result shows that
\begin{enumerate}
\item $m_{\tilde{\nu}} \lt m_W$. Although one can tune the mass of
the sneutrino so the annihilation cross section has a large Z-pole enhancement,
one still needs a large boost factor to have a
sizable excess. Perhaps such a large boost factor is unlikely, but we cannot be sure.

\item $m_{\tilde{\nu}} \ge m_W$. As expected, the result in this case is
very similar to that of the neutralino LSP scenario. It is not identical to the neutralino case because the cross sections are somewhat different.  
\end{enumerate}

Notice that in both cases, there is no ``bump''-like structure resulting
from the LSP annihilations, as expected since we do not
produce anything around $8$ GeV in the first place. The excess is purely
due to the energy loss of the high energy positrons---from W and
fermion decay, including secondary positrons mainly from b, $\mu$ and $\tau$.

\subsection{Phenomenology of This Scenario}
 
The first question is whether the correct amount of relic sneutrinos can
indeed be produced, survive, and constitute the cold dark matter of our
universe. One possibility is standard thermal
production. This always gives far too few relic sneutrinos.  In recent
years,  it has been realized that
non-thermal production mechanisms may dominate, and different
mechanisms\cite{Kallosh:1999jj,Giudice:1999yt,Giudice:1999am,Gherghetta:1999sw,
Jeannerot:1999yn,Moroi:1999zb} of non-thermal 
production have been proposed and studied. Although most of them are in
the context of non-thermal production of neutralinos, they are still
valid in our case since neutralinos will inevitably decay into
sneutrinos. Thus it is possible to produce sufficient sneutrinos by
non-thermal mechanisms, but the incomplete understanding of such
mechanisms means we cannot draw a definite  conclusion.

Next, consider briefly the collider signatures. 
\begin{enumerate}
\item Compatibility with LEP-II results \cite{Heister:2001nk,DELPHY, OPAL, L3}.
For the case of  $m_{\tilde{\nu}} \lt m_W$, we need the
sneutrino mass to be half the Z mass (45.6 GeV) to get the maximum
enhancement. This mass is consistent with the invisible Z width
measurement, which gives a model-independent lower bound of 44.25
GeV. At tree level the associated slepton mass  is determined by
$m_{\tilde{\nu}}^2-m_{\tilde{\it l}_L}^2=m_W^2\cos2\beta$. To be
consistent with LEP slepton results, we need larger $\tan\beta$ to make sleptons
heavy. The lower limit on the Higgs boson mass suggests $\tan \beta \stackrel{>}{_{\sim}}
4$. When $\tan\beta \stackrel{>}{_{\sim}} 4$, $m_{\tilde{\it l}_L}=91.2$ GeV. We must also
take into account loop contributions, so the slepton 
mass will be about 95 GeV\cite{Yamada:1996jf}. This gives a $1 \sigma$ signal expected at
LEP (amusingly, a small excess of order $1 \sigma$ is observed for
smuons by OPAL and DELPHI.) If the LSP is a mixture of left-handed and
right-handed sneutrino, then the mass splitting of slepton and sneutrino
will be larger than the D-term splitting\cite{Arkani-Hamed:2000kj}.

Models with $\tilde\nu$ LSP and $m_{\tilde{\nu}} \ge m_W$ are consistent
with current collider experiments. The $\tilde{\nu}$ could then be
right-handed as well, so long as it can annihilate into W's,  but does not need to be. 
 
\item Tevatron Signals.
At the Tevatron, there can be $\tilde{\it l}_L^\pm\tilde\nu$ and 
$\tilde{\it l}_L^\pm 
\tilde{\it l}_L^\mp$ production. For the first case the signature is a
single charged lepton plus transverse missing energy and
 the cross section for each family is about 246 fb for
$m_{\tilde{\nu}} \lt m_W$  and  
75 fb for $m_{\tilde{\nu}} \ge m_W$.
For the second case the signature is a charged lepton pair with
transverse missing energy and the cross
section for each slepton generation is about 32 fb for  $m_{\tilde{\nu}} \lt
m_W$ and 15 fb for $m_{\tilde{\nu}} \ge m_W$. The light $\tilde{\nu}$
case can surely be observed with $\sim 2 fb^{-1}$ luminosity and
presumably the heavier one with $5-10 fb^{-1}$.  We use PYTHIA for
calculating the cross sections. \cite{Sjostrand:2000wi}

\end{enumerate}

\section{$\tilde{\nu}$, $\tilde{N}_1$ degeneracy}

In this scenario, we made an attempt to generate some structure around
$10$ GeV in a special model. The best thing one can do is to
have a positron production mechanism with a
$\delta$-function distribution centering around $8$ GeV. However, this does not
give a bump but a structure such as that shown in Fig~\ref{delta}, which
is the best possible. Such a source of positrons can only
come from two-body direct production or direct decay 
from a particle that is almost at rest. Consider the decay
first. Since LSP's do not decay, we have to rely on the decay of a Standard
Model particle, and the difference in masses of the decaying
particle and the decay products other than the positron has to be around
$8$ GeV. There is no such particles. Therefore, we have to consider
the possibility of  
production. The production of a pair of fermions will not work since 
there is no decaying particle with a mass of $16$ GeV. The initial state
of the  annihilation must be neutral, and therefore also the final
states. This naturally
leads us to consider the final state $W^- + e^+$. Then R-parity
conservation and lepton number conservation force us to choose the initial
state $\tilde{N}_1+\tilde{\nu}$. 

\subsection{Numerical Results}
The information needed from particle physics in this case are the mass
of he sneutrino/neutralino and the content of the neutralino.  It
is impossible to have a wino/higgsino-like neutralino with the
appropriate mass since it would imply a very light chargino which is not
consistent with the LEP limit. Therefore, we are forced to use a bino-like
lightest neutralino. We also include the boost factor and the
overall normalization of the background positrons in our combined
fit. 

Notice that in this case, sneutrinos and neutralinos also annihilate
among themselves. However, since their masses are less than
$m_W$, those self-annihilation only give a tiny positron excesses. We are
fully relying on the coannihilation to produce the necessary
excess. Therefore, it would be most efficient for the signal if both sparticles
had nearly the same number density. In our calculation here, we assume
the dark matter consists of neutralino and electron sneutrino and their
number densities are equal.  If we assume the 
neutralino and all three families of sneutrino are degenerate, the
number densities for each generation of sneutrino, anti-sneutrino are
equal, and the number densities of neutralino and sneutrino are equal, we
need a boost factor three times larger the one we reported above.

The result is shown in Fig~\ref{spectrum2}. We see that some structure
near $8 - 10$ 
GeV is produced, though it does not resemble the data very well. The main
reason is that 
although we manage to inject a $\delta$-function distribution into the
spectrum before propagation, the one-sided character of energy loss can
not give us a spectrum just like the data.  

\subsection{Phenomenology of $\tilde{\nu}$-$\tilde{N}_1$-degenerate
Scenario}

This model is extremely fine-tuned since we require a degeneracy
of masses $m_{\tilde{\nu}}$ and $m_{\tilde{N}_1}$ to an accuracy of less
than a couple eV. It is of course very unlikely that such an accident can happen
without a symmetry. To the best of our knowledge, a
symmetry that can achieve this is unknown. In extended theories 
it is possible that the sneutrino and neutralino can
appear in the same multiplet, so it is conceivable that such a
symmetry could exist in the underlying theory. However, 
supersymmetry must be broken and the sneutrino and neutralino actually
get mass from the supersymmetry breaking, so it is unlikely such a
symmetry could be preserved. Without data as a motivation we would not consider such a degeneracy. 

Second, it is a more subtle question now if this special
composition of the cold dark matter can be realized. Since the
coannihilation is very effective, the thermal
production will never give us enough relic abundance and  non-thermal
production is required. Mechanisms have been proposed
\cite{Kallosh:1999jj,Giudice:1999yt,Giudice:1999am,Gherghetta:1999sw,Jeannerot:1999yn,
Moroi:1999zb}. 
to produce pairs of neutralinos from the decay of a moduli field. If mainly winos are 
produced, they will decay to binos through $\tilde{W} \rightarrow
\tilde{e}+e \rightarrow \tilde{B} + e + e$ and to sneutrinos through
$\tilde{W} \rightarrow \tilde{\nu} + \nu$, so the relative production of
binos and sneutrinos is about what is needed for the model to be
relevant. This may require a suppression of the 
coupling of the moduli to bino to avoid their overproduction.   
 
Finally, we study the collider signature of this special scenario.
\begin{enumerate}
\item LEP-II.
The constraint on the sneutrino mass is the same as in section 2.2. Here we
have a light neutralino around 50 GeV. If the second lightest neutralino
is also light such that $e^+e^-\to\tilde N_1 \tilde N_2$ is kinematically
allowed, then after $\tilde N_2$ decay, there will be an  acoplanar
lepton pair signal. To suppress this channel, the mass of the second
neutralino should be above 155 GeV. A set of parameters that give a
spectrum which is consistent with all LEP data is 
$M_1=57$ GeV, $M_2=400$ GeV, $\mu=180$ GeV , $\tan\beta=10$. 

\item Tevatron.
The signature of sleptons are the same as in section 2.2. Now we also
have $\tilde C_1\tilde C_1\,,\tilde C_1\tilde N_1\,,\tilde C_1\tilde
N_2$ and $\tilde N_1\tilde N_2$ channels open with cross sections around
41fb, 19fb, 46fb and 0.7fb separately.  The first one and the fourth one,
after $\tilde C_1^\pm$ or $\tilde N_2$ decay, can give a charged lepton
pair plus transverse missing energy. The second one, after $\tilde
C_1^\pm$ decay, can gives single charged lepton with transverse missing
energy. The third one can give a trilepton signal, three charged lepton. 

\end{enumerate}

\section{Detectability by various experiments}
In Table 1. we list detectability by various experiments for our
models. Detectable for the Tevatron means at least one superpartner of
the spectrum containing the LSP would be produced in numbers large
enough to observe. The measured antiproton flux in the evergy region
around 0.5 GeV is about
$1.27\times10^{-6}\mbox{cm}^{-2}\mbox{sr}^{-1}\mbox{sec}^{-1}\mbox{GeV}^{-1} 
$. Estimates for the $\bar p$ flux from
our models are shown in the table. Given the uncertainties we think none
of these are excluded, and all are large enough to see a signal if
qualitatively better measuments can be made\cite{Mori:2001dv}.

\begin{table} {Table 1. Detectability by various experiments. } \\[1ex]
\begin{tabular}{|p{2cm}|p{2cm}|p{3cm}|p{3.3cm}|p{3cm}|} \hline
 Model & Collider & Direct Detector & $\bar p$~flux~({\footnotesize $10^{-6}$
$\mbox{cm}^{-2}\mbox{sr}^{-1}\mbox{sec}^{-1}\mbox{GeV}^{-1}$}) &
underground neutrino
 \\ \hline
 mainly wino $\tilde N$ & Tevatron & detectable soon & $\sim 4.8$ &
detectable soon \\ \hline
 mainly higgsino $\tilde N$ & Tevatron  & no  & $\sim 4.1$
 & detectable soon \\ \hline
 $\tilde\nu$ heavier than W & Tevatron & excluded for simple models &
$\sim 0.7$ &
excluded for simple models \\ \hline
 $\tilde\nu$ lighter than W & Tevatron & excluded for simple models &
$\sim 6.0$ &
excluded for simple models \\ \hline
 degenerate $\tilde\nu$ $\tilde N$ & Tevatron & excluded for simple models
& $\sim 0.8$ & excluded for simple models \\ \hline
\end{tabular}
\end{table}
\section{Conclusions}

We have studied whether positrons from LSP annihilation could account
for the excess and structure in the positron spectrum reported by the HEAT
Collaboration. Even normalizing the relic densities to the local galactic
density, significant ``boost factors'' are sometimes required to get a sufficiently
large signal. It is not known whether such boost factors can be explained
by galactic propagation and local concentrations. Statistically
reasonable descriptions of the data can be obtained for sneutrino or
neutralino LSP's heavier than W bosons, but they lead to a smooth energy
spectrum rather than the apparent peaked structure of the data.  In
simple models, but not generally, sneutrinos  are excluded by the
absence of direct detection. Although no LSP
annihilation approach could give the peaking suggested by the data since
a $\delta$-function input becomes essentially a step function because of
the energy loss, we also construct a fine-tuned model to generate a $\delta$-function positron
energy input, by assuming degenerate sneutrino and neutralino LSP's. The
degenerate LSP model implies the observability of charginos,
neutralinos, and sleptons at the Tevatron, and the sneutrino and
neutralino LSP interpretations both allow detection of signals at the Tevatron.

The description of the data with neutralinos is good, and rather natural
if non-thermal LSP production gives the dominant contribution to the
relic density. The light mass ties in well with indirect evidence for
light superpartners. The sneutrino models seem less attractive, but we
report them because we think assumptions made for theoretical simplicity
should not be taken too seriously when examining potentially important data. 
 
\section{Ackonwledgements}

We appreciate encouragement from and discussions with G.~Tarle,
M.~Schubnell, D.~Chung, and J.~Wells.

\begin{thebibliography}{99}
\bibitem{heat:2000}
S.~Coutu {\it et al.} Proceedings of ICRC 2001. 

\bibitem{Barwick:1997ig}
S.~W.~Barwick {\it et al.}  [HEAT Collaboration],
Astrophys.\ J.\  {\bf 482}, L191 (1997)
[arXiv:astro-ph/9703192].

\bibitem{Barwick:1995gv}
S.~W.~Barwick {\it et al.}  [HEAT Collaboration],
Phys.\ Rev.\ Lett.\  {\bf 75}, 390 (1995)
[arXiv:astro-ph/9505141].

\bibitem{Turner:1989kg}
M.~S.~Turner and F.~Wilczek,
Phys.\ Rev.\ D {\bf 42}, 1001 (1990).

\bibitem{Tylka:xj}
A.~J.~Tylka,
Phys.\ Rev.\ Lett.\  {\bf 63}, 840 (1989)
[Erratum-ibid.\  {\bf 63}, 1658 (1989)].

\bibitem{Kamionkowski:1990ty}
M.~Kamionkowski and M.~S.~Turner,
Phys.\ Rev.\ D {\bf 43}, 1774 (1991).


\bibitem{Baltz:1998xv}
E.~A.~Baltz and J.~Edsjo,
Phys.\ Rev.\ D {\bf 59}, 023511 (1999)
[arXiv:astro-ph/9808243].


\bibitem{Kane:2001fz}
G.~L.~Kane, L.~T.~Wang and J.~D.~Wells,
arXiv:hep-ph/0108138.

\bibitem{Baltz:2001ir}
E.~A.~Baltz, J.~Edsjo, K.~Freese and P.~Gondolo,
arXiv:astro-ph/0109318.

\bibitem{Sikivie:1995dp}
P.~Sikivie, I.~I.~Tkachev and Y.~Wang,
Phys.\ Rev.\ Lett.\  {\bf 75}, 2911 (1995)
[arXiv:astro-ph/9504052].

\bibitem{Gondolo:2000ee}
P.~Gondolo, J.~Edsjo, L.~Bergstrom, P.~Ullio and E.~A.~Baltz,
``DarkSUSY: A numerical package for dark matter calculations in the  MSSM,''
arXiv:astro-ph/0012234.

\bibitem{Moskalenko:1997gh}
I.~V.~Moskalenko and A.~W.~Strong,
Astrophys.\ J.\  {\bf 493}, 694 (1998)
[arXiv:astro-ph/9710124].

\bibitem{Hall:1997ah}
L.~J.~Hall, T.~Moroi and H.~Murayama,
Phys.\ Lett.\ B {\bf 424}, 305 (1998)
[arXiv:hep-ph/9712515].

\bibitem{Arkani-Hamed:2000kj}
N.~Arkani-Hamed, L.~J.~Hall, H.~Murayama, D.~R.~Smith and N.~Weiner,
arXiv:hep-ph/0007001.


\bibitem{Smith:2001hy}
D.~R.~Smith and N.~Weiner,
Phys.\ Rev.\ D {\bf 64}, 043502 (2001)
[arXiv:hep-ph/0101138].

\bibitem{Kallosh:1999jj}
R.~Kallosh, L.~Kofman, A.~D.~Linde and A.~Van Proeyen,
Phys.\ Rev.\ D {\bf 61}, 103503 (2000)
[arXiv:hep-th/9907124].

\bibitem{Giudice:1999yt}
G.~F.~Giudice, I.~Tkachev and A.~Riotto,
JHEP {\bf 9908}, 009 (1999)
[arXiv:hep-ph/9907510].

\bibitem{Giudice:1999am}
G.~F.~Giudice, A.~Riotto and I.~Tkachev,
JHEP {\bf 9911}, 036 (1999)
[arXiv:hep-ph/9911302].

\bibitem{Gherghetta:1999sw}
T.~Gherghetta, G.~F.~Giudice and J.~D.~Wells,
Nucl.\ Phys.\ B {\bf 559}, 27 (1999)
[arXiv:hep-ph/9904378].

\bibitem{Jeannerot:1999yn}
R.~Jeannerot, X.~Zhang and R.~H.~Brandenberger,
JHEP {\bf 9912}, 003 (1999)
[arXiv:hep-ph/9901357].

\bibitem{Moroi:1999zb}
T.~Moroi and L.~Randall,
Nucl.\ Phys.\ B {\bf 570}, 455 (2000)
[arXiv:hep-ph/9906527].


\bibitem{Orito:1999re}
S.~Orito {\it et al.}  [BESS Collaboration],
Phys.\ Rev.\ Lett.\  {\bf 84}, 1078 (2000)
[arXiv:astro-ph/9906426].

\bibitem{Maeno:2000qx}
T.~Maeno {\it et al.}  [BESS Collaboration],
Astropart.\ Phys.\  {\bf 16}, 121 (2001)
[arXiv:astro-ph/0010381].

\bibitem{Beach:2001ub}
A.~S.~Beach {\it et al.},
Phys.\ Rev.\ Lett.\  {\bf 87}, 271101 (2001)
[arXiv:astro-ph/0111094].


\bibitem{Altarelli:2001wx}
G.~Altarelli, F.~Caravaglios, G.~F.~Giudice, P.~Gambino and G.~Ridolfi,
JHEP {\bf 0106}, 018 (2001)
[arXiv:hep-ph/0106029].

\bibitem{Cho:1999km}
G.~C.~Cho and K.~Hagiwara,
Nucl.\ Phys.\ B {\bf 574}, 623 (2000)
[arXiv:hep-ph/9912260].

\bibitem{Pukhov:1999gg}
A.~Pukhov {\it et al.},
arXiv:hep-ph/9908288.



\bibitem{Heister:2001nk}
A.~Heister {\it et al.}  [ALEPH Collaboration],
Phys.\ Lett.\ B {\bf 526}, 206 (2002)
[arXiv:hep-ex/0112011].



\bibitem{DELPHY} DELPHI Coll., ``Searches for supersymmetric particles
in $e^+e^-$ collisions up to 208 GeV, and interpretation of the results
within the MSSM'', DELPHI 2001-085 CONF 513.

\bibitem{L3} The L3 collaboration, ``Search for Supersymmetry in
$e^+e^-$ collisons at $\sqrt s=200-208$GeV, L3 note 2707

\bibitem{OPAL} OPAL Coll. ``New Particle Searches in $e^+e^-$ Collisons
at $\sqrt{s}=200-209$GeV'', OPAL Physics Note PN470.




\bibitem{Yamada:1996jf}
Y.~Yamada,
Phys.\ Rev.\ D {\bf 54}, 1150 (1996)
[arXiv:hep-ph/9602279].




\bibitem{Sjostrand:2000wi}
T.~Sjostrand, P.~Eden, C.~Friberg, L.~Lonnblad, G.~Miu, S.~Mrenna and E.~Norrbin,
Comput.\ Phys.\ Commun.\  {\bf 135}, 238 (2001)
[arXiv:hep-ph/0010017].

\bibitem{Mori:2001dv}
K.~Mori, C.~J.~Hailey, E.~A.~Baltz, W.~W.~Craig, M.~Kamionkowski, W.~T.~Serber and P.~Ullio,
arXiv:astro-ph/0109463.


\end{thebibliography}
\end{document}